\documentclass[reprint,superscriptaddress,nofootinbib,amsmath,amssymb,aps,prl,onecolumn,notitlepage]{revtex4-1}

\usepackage[a4paper, margin=2.50cm]{geometry}

\usepackage{bbm}
\usepackage{graphicx}
\usepackage{fancyhdr, color}
\usepackage{hyperref}

\newcommand{\ket}[1]{\left| #1 \right\rangle}
\newcommand{\bra}[1]{\left\langle #1 \right|}

\newcommand{\avg}[1]{\left\langle #1 \right\rangle}

\begin{document}

\fancyhead[C]{\sc \color[rgb]{0.4,0.2,0.9}{Quantum Thermodynamics book}}
\fancyhead[R]{}

\title{Probing Quantum Fluctuations of Work with a Trapped Ion}

\author{Yao Lu}
\email{luyao_physics@163.com} 
\affiliation{Center for Quantum Information, Institute for Interdisciplinary Information Sciences, Tsinghua University, Beijing 100084, P. R. China}

\author{Shuoming An}
\email{anshuoming-hn@163.com} 
\affiliation{Center for Quantum Information, Institute for Interdisciplinary Information Sciences, Tsinghua University, Beijing 100084, P. R. China}

\author{Jing-Ning Zhang}
\email{jnzhang13@mail.tsinghua.edu.cn} 
\affiliation{Center for Quantum Information, Institute for Interdisciplinary Information Sciences, Tsinghua University, Beijing 100084, P. R. China}

\author{Kihwan Kim}
\email{kimkihwan@mail.tsinghua.edu.cn} 
\affiliation{Center for Quantum Information, Institute for Interdisciplinary Information Sciences, Tsinghua University, Beijing 100084, P. R. China}

\date{\today}

\begin{abstract}
In this chapter, we illustrate how a trapped ion system can be used for the experimental study of quantum thermodynamics, in particular, quantum fluctuation of work. As technology of nano/micro scale develops, it becomes critical to understand thermodynamics at the quantum mechanical level. The trapped ion system is a representative physical platform to experimentally demonstrate quantum phenomena with excellent control and precision. We provide a basic introduction of the trapped ion system and present the theoretical framework for the experimental study of quantum thermodynamics. Then we bring out two concrete examples of the experimental demonstrations. Finally, we discuss the results and the future of the experimental study of quantum thermodynamics with trapped ion systems.  

\end{abstract}

\maketitle

\thispagestyle{fancy}

\section{Introduction} \label{sec:introduction}
The technologies of nano/micro scale have been rapidly developed~\cite{Szabo01,Liphardt2002,Collin2005}. On the nano/micro scale, quantum fluctuations become important and such quantum effects should be fully taken into account to properly understand and manipulate nano/micro systems~\cite{Szabo01,Liphardt2002,Collin2005,Douarche2005,Blickle2006,Harris2007, Junier2009, Shank2010, Saira2012}. The extension of the principles of thermodynamics~\cite{Jarzynski1997a,Crooks99} to the quantum regime has been the subject of extensive theoretical studies in the last decades~\cite{Kurchan2000, Tasaki2000, Mukamel2003, Talkner2011, Kehrein12, Dorner13, Mazzola13, Hanggi2015}. As nano/micro technologies develop, there has been growing interest in experimentally testing the principles of quantum thermodynamics, which would be a solid foundation for further development of technologies in the quantum regime. 

Among diverse physical platforms, a trapped ion system is particularly interesting to test quantum thermodynamics~\cite{Huber08,HuberThesis,An15,An16,Smith18}. One important feature in a quantum system, which distinguishes from classical macroscopic systems, is energy quantization. Historically, the finding of discretized energy levels of an atom has led to the birth of quantum mechanics~\cite{Dirac} and the electron shelving technique can distinguish internal states of the trapped ion with near-perfect detection efficiency~\cite{Dehmelt75, Bergquist86,Nagourney86,Sauter86,Blatt88}. For the external degree of freedom of the atomic ion trapped in a harmonic potential, the technology of cooling to the ground state of the trap has been developed and the quantized energy levels of motion have been observed and manipulated through the resolved Raman laser technique~\cite{Wineland75a, Wineland78, Neuhauser78, Diedrich89, Leibfried03, Haffner08}. The single trapped atomic ion can be considered as the most fundamental and the simplest system, which can be manipulated and measured with high precision enough to experimentally verify the principles of quantum thermodynamics. 

Generally, the principles of  non-equilibrium thermodynamics are expressed as inequalities. For example, the work performed on a system in contact with a heat reservoir during an arbitrary drive of the system is lower bounded by the net change in its free energy: $\avg{W} \geq \Delta F$, where the equality holds for equilibrium processes. When statistical fluctuations are properly incorporated in exponential form, the inequality is reformulated as equality, which is known as the Jarzynski equality~\cite{Jarzynski1997a}, 
\begin{equation}
\langle e^{-W/k_{\rm B}T}\rangle= e^{-\Delta F/k_{\rm B}T},\label{eq:JarzynskiEq}
\end{equation}
where $T$ is the temperature of the environment and the brackets $\avg{\cdot}$ denote an average over repetitions of the process. For classical systems, this prediction and related fluctuation theorems have been extensively studied both theoretically~\cite{Chipot2007,Pohorille2010,Jarzynski2011} and experimentally~\cite{Liphardt2002, Collin2005, Douarche2005, Blickle2006, Harris2007, Junier2009, Shank2010, Saira2012}. 

However, it has been shown that extending these advances to quantum systems is challenging. One main difficulty is to measure work and work distributions in a quantum system~\cite{Hanggi07}. In classical systems, work can be obtained by measuring work trajectories, which cannot be determined in quantum regime. For closed quantum systems with no heat transfer to or from the system, the difficulty can be avoided and the work can be obtained by measuring the difference of the initial and the final energies~\cite{Kurchan2000, Tasaki2000, Mukamel2003}, which has been demonstrated experimentally~\cite{Batalhao2014,An15}. However, it still remains as the challenge to extend the equality and the theorem to open quantum systems~\cite{Esposito2009}. Recently, for the restricted open environment that induces only dephasing not dissipation, it was pointed out that the equality would hold in quantum regime~\cite{Smith18}.  

In this chapter, we present the experimental tests of the Jarzynski equality for both closed and quasi-open quantum systems based on trapped ion systems~\cite{An15,Smith18}. In the following section, we provide a basic introduction of the trapped ion system, then present the theoretical framework for the experimental test. Finally we discuss the examples of experimental tests of the Jarzynski equality and conclude the chapter. 

\section{Trapped-Ion System} \label{sec:trapped_ion_systems}

\subsection{Free Hamiltonian} 
\begin{figure*}[ht]
\centering
\includegraphics[width=0.8\textwidth]{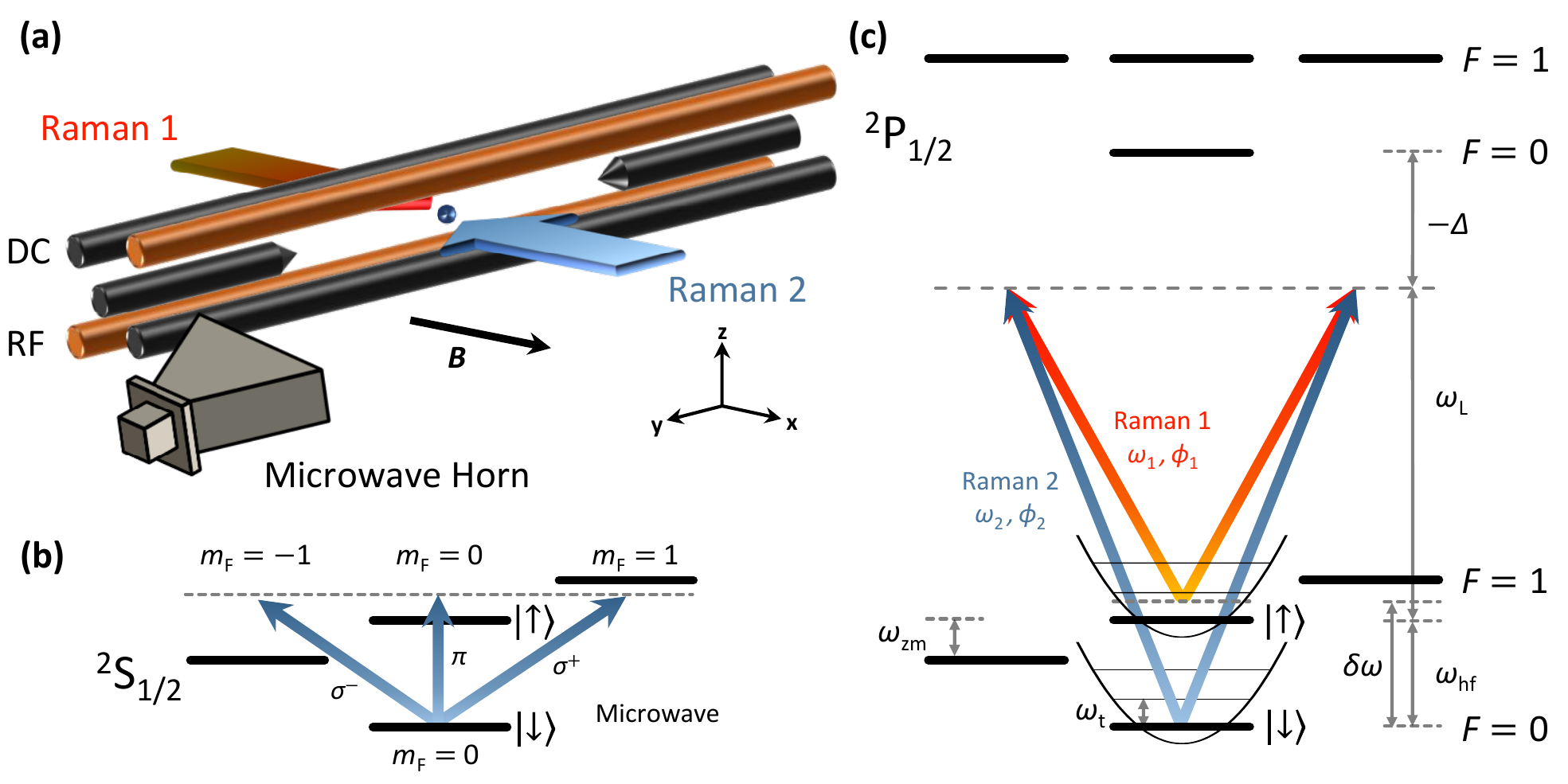}
\caption{Trapping and manipulation of the $^{171}\mathrm{Yb}^+$ ion. (a) Schematic plot of a four-rod trap. The ion is trapped at the center of the four-rod trap, and manipulated either by microwave from the microwave horn or by a pair of counter-propagating Raman beams along the $X$-direction. The external magnetic field is applied along the $X$ direction. (b) Microwave manipulation and relevant energy level diagram. The internal states in the ground state $^2 S_{1/2}$ manifold are manipulated by the microwave radiation. To target transitions between different states, we first introduce an external magnetic field to induce the linear Zeeman splitting $\omega_{\rm zm}$ of the $F=1$ manifold, and then tune the frequency of the microwave to the desired resonance. (c) Raman manipulation and relevant energy level diagram. The stimulated Raman process, which is implemented by two laser beams coupling to the $^2P_{1/2}$ states with a large detuning $-\Delta$, has the ability to couple the internal and external degrees of freedom. Thus the lower part of the energy spectrum can be viewed as a harmonic oscillator superposed on a two-level system. By tuning the frequency difference $\delta\omega$ of the two Raman beams around the hyperfine splitting $\omega_{\rm hf} = 2\pi\times12.6{\rm GHz}$, we can selectively excite or dexcite the motional states when manipulating the internal states.
}
\label{fig:Setup}
\end{figure*}

As one of the most promising platforms for quantum simulation~~\cite{Zhang17,Zhang17_2} and quantum computation~~\cite{Debnath16,Monz16}, the trapped-ion system is famous for its unprecedented stability and controllability. A typical trapped-ion system is shown in Fig.$\,$\ref{fig:Setup}(a), where a single atomic ion is confined in a harmonic potential produced by oscillating radio-frequency electric field~~\cite{Paul90,Leibfried03}. The state of the trapped-ion system belongs to the product of the internal and the external Hilbert spaces. The internal space is expanded by  a selection of electronic energy levels. In our setup with $^{171}\mathrm{Yb}^+$, the basis of the internal space is defined by a pair of hyperfine states of the ground state manifold $^2 S_{1/2}$. For instance $\ket{F=0,\,m_F=0}=\ket{\downarrow}$ and $\ket{F=1,\,m_F=0}=\ket{\uparrow}$ shown in Figs.$\,$\ref{fig:Setup}(b,c), known as the clock states, are a good choice of basis because of the long coherence time~~\cite{Wang17}. Sometimes the upper state $\ket{\uparrow}$ is encoded with $\ket{F=1,\,m_F= \pm 1}$ for a certain purpose. Note that the internal state of a single trapped ion can be represented by the Pauli matrices ${\hat{\sigma}_{\rm x,y,z}}$ and the identity $\hat{\mathbb I}$, known as the Bloch representation.

The other part of the Hilbert space is related to the motion of the ion, which can be modeled by a harmonic oscillator. In this case, a set of the Fock number states $\{|n\rangle, n = 0,1,2 \cdots\}$, also known as the phonon number states, could be chosen as a complete orthonormal basis of this space. The free Hamiltonian of a single trapped-ion system can be written as 
\begin{equation}
\hat{H}_0 = \dfrac{\hbar\omega_0}{2} \hat{\sigma}_\mathrm{z} + \hbar\omega_{\rm t} \left(\hat{a}^\dagger\hat{a} + \dfrac{1}{2}\right),
\end{equation}
where $\omega_0$ is the internal energy splitting, and $\hat{a}$ ($\hat{a}^\dagger$) is the annihilation (creation) operator for the harmonic oscillator with the trapping frequency $\omega_{\rm t}$. As shown in Figs.$\,$\ref{fig:Setup}(b,c), depending on the choice of the electronic states to encode the two-level system, the internal energy splitting is $\omega_0 = \omega_{\rm hf} +\Delta_{m_{\rm F}} \omega_{\rm zm}$, where $\omega_{\rm hf}$ is the hyperfine splitting, $\Delta_{m_{\rm F}}=0, \pm1$ is the difference of the magnetic quantum numbers, and $\omega_{\rm zm}$ is the first-order Zeeman splitting of the $F=1$ hyperfine manifold.

\subsection{Quantum Manipulation}

In order to manipulate the quantum state of the trapped-ion system, either microwaves or laser beams should be applied to the $^{171}\mathrm{Yb}^+$ ion. For the magnetic dipole interaction induced by the microwave shown in Fig.$\,$\ref{fig:Setup} (b), the interaction Hamiltonian can be written as
\begin{equation}
\hat{H}_\mathrm{i}^\mu = \hbar\Omega \hat\sigma_\mathrm{x} \cos(\omega t - \phi),
\end{equation}
where $\Omega$ is the Rabi oscillation frequency, $\omega$ and $\phi$ are the frequency and the initial phase of the applied microwave, respectively. Using the time evolution operator $\hat{U}_0= \exp [-(i/\hbar) \hat{H}_0 t]$, $\hat{H}_\mathrm{i}^{\mu}$ can be transformed into the interaction picture
\begin{align}
\hat{H}_{\mathrm{I}}^{\mu} &= \hat{U}^\dagger_0 \hat{H}_\mathrm{i}^{\mu} \hat{U}_0 \nonumber\\
&= \dfrac{\hbar\Omega}{2}[\hat{\sigma}_\mathrm{x} \cos(\delta t - \phi) + \hat{\sigma}_\mathrm{y} \sin(\delta t - \phi)],
\end{align}
where $\delta = \omega - \omega_0$ is the frequency difference between the microwave and the energy splitting of the internal two-level system.

The electric dipole interaction induced by light fields is relatively complicated because the coupling to the motional degrees of freedom cannot be neglected. The Hamiltonian is written as follows, 
\begin{equation}
\hat{H}^{\rm L}_\mathrm{i} = \hbar\Omega \hat\sigma_\mathrm{x} \cos\left(k\hat x - \omega t + \phi\right), \label{eq:laser-ion}
\end{equation}
where $k$ is the effective component of the momentum along the $X$-axis and $\hat x=\sqrt{\frac{\hbar}{2M\omega_{\rm t}}}\left(\hat a+\hat a^\dag\right)$ is the $X$-component of the position operator of the ion and $M$ is the mass of the single $^{171}\mathrm{Yb}^+$ ion. In the $^{171}\mathrm{Yb}^+$ system, we off-resonantly couple the $\ket{\uparrow}$ and the $\ket{\downarrow}$ states to the excited $^2 P$ manifold, constructing a $\Lambda$-type Raman transition as shown in Fig.$\,$\ref{fig:Setup} (c), to realize the laser-ion interaction described by Eq.$\,$(\ref{eq:laser-ion}). By tuning the frequency difference $\omega = \omega_2 -\omega_1$ (assuming $\omega_2 > \omega_1$) of two Raman lasers, several kinds of the resonant transitions can be realized in the interaction picture, for example,
\begin{itemize}
\item The carrier transition with $\omega = \omega_0$,
\begin{eqnarray}
\hat{H}_\mathrm{c}= \dfrac{\hbar\Omega_{\mathrm{eff}}}{2} (\hat\sigma_{+} e^{i \phi} + \hat\sigma_{-} e^{- i \phi}).
\end{eqnarray}
\item The red-sideband transition with $\omega = \omega_0 - \omega_{\rm t}$,
\begin{eqnarray}\label{eq:rsb}
\hat{H}_\mathrm{rsb} = \dfrac{\hbar\eta\Omega_\mathrm{eff}}{2} (\hat\sigma_{+} \hat{a} e^{i \phi} + \hat\sigma_{-} \hat{a}^\dagger e^{- i \phi}).
\end{eqnarray}
\item The blue-sideband transition with $\omega = \omega_0 + \omega_{\rm t}$,
\begin{eqnarray}\label{eq:bsb}
\hat{H}_\mathrm{bsb} = \dfrac{\hbar\eta\Omega_\mathrm{eff}}{2} (\hat\sigma_{+} \hat{a}^\dagger e^{i \phi} + \hat\sigma_{-} \hat{a} e^{- i \phi}).
\end{eqnarray}
\end{itemize}

Here, $\hat\sigma_{\pm}$ is spin raising (lowering) operator, $\eta = k\sqrt{\hbar/(2M\omega_{\rm t})}$ is the Lamb-Dicke parameter with $k$ being the component of the momentum difference of the two Raman beams along the $X$-axis, while $\phi = \phi_2 - \phi_1$ and $\Omega_\mathrm{eff}$ are the relative phase between two light fields and the effective coupling strength of the Raman transition, respectively.

\subsection{Initialization and Measurement}

\subsubsection{State Initialization}
The quantum state initialization is the first step for typical experimental protocols with the trapped ion system. Depending on the requirement of Hilbert space in the protocols, we initialize only the internal state or both of the internal and the motional state. The initialization of the internal state is performed by utilizing optical pumping. In the $^{171}{\rm Yb}^{+}$ ion, the optical transition between $\ket{^2S_{1/2}\, ,F=1}$ and $\ket{^2P_{1/2}\, ,F=1}$ leaves the population in the $\ket{\downarrow}$ state~\cite{Olmschenk07}.

As to the motional state, we introduce a two-stage cooling protocol to prepare the ground state of the harmonic oscillator. We first perform the standard Doppler cooling through the transition from $^2S_{1/2}$ to $^2P_{1/2}$ with a red detuning, and cool the motion of the ion to an average phonon number of 15. To further lower the ion's temperature, we continue to perform the resolved sideband cooling. Specifically, we alternately apply the red sideband transition and the optical pumping sequences for around hundred times. The final state has an average phonon number of around 0.02, which is the ground state within experimental precision.

\subsubsection{Measurement of State}

The quantum measurement for trapped-ion systems includes the measurements of the internal states and the motional states. The internal states are measured by the state-dependent fluorescence detection~~\cite{Olmschenk07}. For $^{171}{\rm Yb}^{+}$, the detection of the internal state is realized by applying a resonant laser to drive the cycling transition between $\ket{^2S_{1/2}\,,F=1}$ and $\ket{^2P_{1/2}\,,F=0}$ for a certain duration and counting the fluorescent photons during this time. Almost no photons are detected if the internal state collapses to the down state $\ket{\downarrow}$, while lots of photons are detected if the upper state $\ket{\uparrow}$ is projected. We repeat the sequence of the state detection for a certain number of times and measure $P_{\uparrow}$, the probability of being in the $\ket{\uparrow}$ state, as the ratio of the number of the fluorescent cases to the total number of repetitions.

For the motional state detection, we map the information of a motional state to the internal state~~\cite{Leibfried03}. By driving the resonant blue-sideband transition on a motional state $\ket{\downarrow}\bigotimes\sum_{n=0}^{\infty}c_n\ket{n}$ for varied interaction duration $\tau$, we measure the upper state probability, 
\begin{align}
P_\uparrow(\tau) = \dfrac{1}{2}[1-\sum_{n=0}^{\infty}|c_n|^2 \cos{(\sqrt{n+1} \Omega_{\rm eff} \tau)}],
\label{eq:bluedetection}
\end{align}
which oscillates with the duration of the blue-sideband transition $\tau$. By applying the Fourier transform to the output signal, we can resolve the phonon distributions $\{|c_n|^2\,,n=0,1,2,3...\}$. We note that for the test of quantum fluctuation theorems, the above ensemble measurement of motional degrees of freedom is not sufficient. We need the projective measurement of phonons to determine the phonon number in each sequence, which will be discussed later.

\section{Experimental framework} \label{sec:model_hamiltonians}

We have shown that the trapped-ion systems are of great controllability, in a sense that state preparation, versatile operations, and state measurements be reliably performed in this platform. Thus the trapped-ion platform is an excellent experimental candidate for exploiting and verifying the foundations of the interdisciplinary field of thermodynamics and quantum mechanics, i.e. quantum thermodynamics.

Internal energy, work and heat belong to the set of the most fundamental concepts in classical thermodynamics, which are related by the first law of thermodynamics. However, the generalization of these classical concepts to the quantum region is not straightforward, where the state of the system is in general a coherent superposition, instead of a mixture, of the energy eigenstates. What's more, these energy eigenstates form a discrete energy spectrum in the deep quantum realm. These intrinsic differences render rich and exotic insights in quantum thermodynamic processes, which have attracted great interest in recent years.

As a starting point, we begin by considering closed and quasi-open quantum systems. In the former case, with the exception of a time-dependent parameter to model the external control, the system does not interact with the environment. Meanwhile, in the latter cases, the system is interacting with a special type of environment, i.e. a dephasing environment, which only eliminates quantum coherence with respect to the energy eingenbasis. In this case, it can be strictly proved that there is no heat transfer during the process~\cite{Smith18}.

In the following, we first introduce the effective models which are feasible in the trapped-ion system. Then we investigate the concept of quantum work in our limited configurations, from both theoretic and experimental perspectives.

\subsection{Effective Models in Rotating Frames}

Recall that by irradiation of electromagnetic fields with suitable frequencies, two of the internal states and/or one of the external modes can be coherently coupled to each other. Here in this section, we show that the parameters of the effective models can be controlled to evolve according to designed time-dependent functions by tuning the intensity and the relative phase of the Raman laser beams. In this manner, we are able to simulate far-from-equilibrium thermodynamic processes in properly chosen rotating frames. Specifically, we deal with two kinds of effective models: 1) the dragged quantum harmonic oscillator and 2) the driven quantum two-level system with the dephasing bath. 

\subsubsection{The Rotating Frame}

A natural and widely adopted rotating frame in a trapped ion system is with respect to the free Hamiltonian $\hat H_0$. In order to gain more flexibility, we introduce a different rotating frame with respect to 
\begin{eqnarray}\label{eq:h_0p}
\hat H_0'(\omega_{\rm z},\nu) =\frac{\hbar\left(\omega_0-\omega_{\rm z}\right)}{2}\hat\sigma_{\rm z}+\hbar\left(\omega_{\rm t}-\nu\right)\left(\hat a^\dag\hat a+\frac{1}{2}\right),
\end{eqnarray}
where the interaction-picture Hamiltonian becomes 
\begin{eqnarray}
\hat H_{\rm I}=\frac{\hbar\omega_{\rm z}}{2}\hat\sigma_{\rm z}+\hbar\nu\left(\hat a^\dag\hat a+\frac{1}{2}\right).
\end{eqnarray}
It is worth bearing in mind that $\omega_{\rm z}$ and $\nu$ are completely determined by the choice of the rotating frame, which is a purely mathematical operation.

\subsubsection{Dragged Quantum Harmonic Oscillators}

The key ingredient needed to construct the model Hamiltonian of a dragged quantum harmonic oscillator is the interaction between the trapped ion and a bichromatic laser field with detuned first red- and blue-sideband frequencies $\omega_\pm=\omega_0\pm\left(\omega_{\rm t}-\nu\right)$. In the rotating frame with respect to $\hat H_0'(0,\nu)$ and after the rotating-wave approximation (RWA), the interaction-picture Hamiltonian is
\begin{eqnarray}\label{eq:hamil_blue_red}
\hat H_{\rm I}(t) = \hbar\nu\left(\hat a^\dag\hat a+\frac{1}{2}\right) +\frac{\hbar\Omega(t)}{2}\left(\hat a+\hat a^\dag\right)\hat\sigma_{\rm x},
\end{eqnarray}
where the Rabi coupling strength $\Omega(t)\equiv \eta\Omega_\pm(t)$ can be varied by tuning the intensity of the bichromatic laser field.

Inspecting the Hamiltonian in Eq.~(\ref{eq:hamil_blue_red}), one can easily find that by setting the initial state of the two-level system to the eigenstate of $\hat\sigma_{\rm x}$ with the eigenvalue $+1$, the two-level system can be factored out and the system reduced to a dragged quantum harmonic oscillator. Rewritten in the first quantized operators, the time-dependent system Hamiltonian becomes
\begin{eqnarray}
\hat H_{\rm S}(t)=\frac{\hat P^2}{2M_{\rm e}}+\frac{1}{2}M_{\rm e}\nu^2\hat X^2+f(t)\hat X,
\label{eq:DH}
\end{eqnarray}
where the momentum operator $\hat P$ and the position operator $\hat X$ are defined as
\begin{eqnarray}
\hat P&=&i\sqrt{\frac{\hbar M_{\rm e}\nu}{2}}\left(\hat a^\dag-\hat a\right),\\
\hat X&=&\sqrt{\frac{\hbar}{2M_{\rm e}\nu}}\left(\hat a^\dag+\hat a\right),\nonumber
\end{eqnarray}
with the effective mass $M_e=\frac{\omega_{\rm t}}{\nu}M$ and the effective dragging force $f(t)=\frac{\hbar}{2}k\Omega(t)$.

Here we have shown how to generate the effective Hamiltonian of a dragged quantum harmonic oscillator by the ion-laser interaction. Furthermore, with more complicated setting of lasers, it is also possible to realize the effective Hamiltonian of a quantum harmonic oscillator with time-dependent frequency in the rotating frame in a pretty similar manner~\cite{Funo17}.

\subsubsection{Driven Quantum Two-Level Systems}

Alternatively, we can use the two-level system to explore the quantum fluctuation theorems. Compared to the dragged harmonic oscillator case, the advantage of this scenario is that it allows us to investigate the effect of the coupling to a special kind of environment, namely the dephasing environment.
 
In order to simulate the quantum dynamics of a driven quantum two-level system, we irradiate the trapped ion by a microwave field with the resonant carrier frequency $\omega_0$. Note that the Lamb-Dicke parameter, which couples the internal and motional degrees of freedom, of a microwave is negligibly small,  thus the motional mode is irrelevant and can be omitted in this case. The effective Hamiltonian in the rotating frame with respect to $\hat H_0$ is written as follows,
\begin{eqnarray}
\hat H_{\rm S}\left(t\right)=\frac{\hbar\Omega(t)}{2}\left[\hat\sigma_{\rm x}\cos\phi(t)+\hat\sigma_{\rm y}\sin\phi(t)\right]\label{eq:TLS_drivenH},
\end{eqnarray}
where the Rabi coupling $\Omega(t)$ and phase $\phi(t)$ can be varied by tuning the intensity and phase of the microwave field.

\subsubsection{Dephasing Environment}

The general coupling to a thermal environment results in non-unitary evolution of the system, whose effect can be phenomenologically classified into two kinds of decohering effects: dissipation and dephasing. The former induces energy transfer between the system and the environment, while the latter only attenuates the quantum coherence of the state, which is quantified by the off-diagonal matrix element of the density operator in the energy eigenbasis. Note that the term ``decoherence'' in Ref.~\cite{Smith18} is used with the same meaning as the term ``dephasing'' here. Although it is possible to have both decohering effects in the trapped-ion system, we only focus on the dephasing effect, because there is no energy transfer between the system and the environment and the definition of work through energy difference is still valid in this case.

With the existence of a purely dephasing environment, the quantum dynamics of the system is described by the following master equation,
\begin{eqnarray}
\frac{d\hat\rho}{dt}=-\frac{i}{\hbar}\left[\hat H_{\rm S}(t),\hat\rho\right]-\sum_{i\neq j}\gamma_{ij}\rho_{ij}\ket{i}\bra{j},
\label{eq:dephasing}
\end{eqnarray}
where $\rho$ is the density operator of the system, $\{\ket{i}\}$ is the instantaneous energy eigenbasis of $\hat H_{\rm S}(t)$, and $\rho_{ij}$ and $\gamma_{ij}$ are the matrix element of $\hat\rho$ in the instantaneous energy eigenbasis and corresponding phenomenological dephasing strength.

In the trapped-ion setup, we introduce the pure dephasing with respect to the instantaneous energy eigenbasis by adding Gaussian white noise into the intensity of the microwave field. The net effect of this operation is the substitution $\Omega(t)\rightarrow \Omega(t)+\Omega_0\xi(t)$, with $\Omega_0\equiv\Omega(t=0)$ and $\xi(t)$ satisfying
\begin{eqnarray}
\avg{\xi(t)}=0,\quad\avg{\xi(t)\xi(t+\tau)}= \sigma^2\delta(\tau),\label{eq:noise}
\end{eqnarray}
with $\sigma$ characterizing the standard deviation of the noise fluctuations, which can be controlled by changing the intensity of the Gaussian white noise and is related to the strength of the dephasing effect~\cite{Loreti94,VKampen2007}.

\subsection{Quantum Work}

\subsubsection{Definition of Quantum Work}
The definition of work in the quantum regime is tricky~\cite{Talkner2011}. In classical systems, work can be obtained by measuring the force and the displacement, and then integrating the force over the displacement during the driving process. In the quantum regime, however, as a result of Heisenberg’s uncertainty principle, we cannot determine the position and the momentum simultaneously thus invalidating the above approach.

We mainly consider the situations of the closed quantum system or the quasi-open quantum system undergoing dephasing. Because of the first law of thermodynamics, work equals the difference of the internal energy and the heat transfered between the system and the environment, namely
\begin{eqnarray}
W = \Delta U-Q.
\end{eqnarray}
Since there is no heat in our consideration, work can be defined as the difference of internal energy before and after applying the work.  
The quantum work is defined as
\begin{equation}
W_{n\rightarrow m}=\epsilon_{\rm f}^{m}-\epsilon_{\rm i}^{n},\label{QW}
\end{equation}
where $\epsilon_{\rm f}^{m}$ is the $m$-th eigenenergy of the final system and $\epsilon_{\rm i}^{n}$ is the $n$-th eigenenergy of the initial system. 

\subsubsection{Measurement of Quautum Work: Two-Point Measurement}
The quantum work in a single realization defined as Eq.~(\ref{QW}) can be obtained by performing the projective measurements over the energy eigenstates at the initial and final points of the process of applying work. We repeat the two-point projective measurements to determine the work distribution~\cite{talkner2007fluctuation,Esposito2009}. In the trapped ion system, we choose to use either the spin or the motional states as the system that we apply the work on. For the spin system, we apply the $\hat{\sigma}_{\rm z}$ measurement along the instantaneous energy eigenbasis, which is equivalent to the energy measurement. For the motional or phonon system, we realize the phonon projective measurement for the first time to get the quantum work $W_{n\to m}$ and its corresponding probability $P_{ n\to m}$. 

The system Hamiltonians in general do not commute with the projective measurements. Intuitively, this can be tackled by adiabatically transferring the probability distribution on the energy basis of the system Hamiltonians to the distribution on the basis of the projective measurements, which commutes with that of the lab frame. Practically, we introduce the technique referred to as the shortcut to adiabaticity to speed up the adiabatic process~\cite{Berry09}.

\section{Examples: The Quantum Jarzynski Equality} \label{sec:examples}
The Jarzynski equality relates the free-energy difference $\Delta F$ to the exponential average of the work $W$ done on the system~\cite{Jarzynski1997a} as shown in Eq.~(\ref{eq:JarzynskiEq}). Its quantum version is introduced later in Refs.~\cite{Esposito2009}. In the trapped ion system, we use the motional state of the ion to test the Jarzynski equality in the closed quantum system and the two-level system consisting of internal states to test that in equality the quasi-open quantum system with the dephasing effect. 

The general experiment sequence for a single realization is summarized as~\cite{Huber08}:
\begin{enumerate}
	\item Prepare the initial thermal state at temperature $T$ with respect to $\hat H_{\rm S}(0)=\sum_n\epsilon_{\rm i}^n\ket{n}\bra{n}_i$,
    \begin{eqnarray*}    
    \hat\rho(T) &=& \exp\left[-\frac{\hat H_{\rm S}(0)}{k_{\rm 		B}T}\right]/{\mathcal Z_0}\\
    &=&\sum_nP_n^{\rm th}\ket{n}\bra{n}_{\rm i}
    \end{eqnarray*}
    with the initial distribution probability $P_n^{\rm th}=\exp\left[-	\frac{\epsilon_{\rm i}^n}{k_B T}\right]/{\mathcal Z_0}$ and ${\mathcal Z_0}=\sum_n\exp\left[-\frac{\epsilon_{\rm i}^n}{k_B T}\right]$ being the initial partition function.
	\item Project to the energy eigenstate $|n\rangle_{\rm i}$ with the probability $P^{\rm th}_{ n}$, 
	\item Apply work $W$ on the system,
	\item Project to the energy eigenstate $|m\rangle_{\rm f}$ of the final state with the probability $P_{ n\to m}$.
\end{enumerate}
The whole sequence is repeated many times to obtain statistics and we can construct the distribution of the work and the exponential average of the work as 
\begin{equation}
\langle e^{-W/k_{\rm B}T}\rangle=\sum P^{\rm th}_{ n}P_{ n\to m}e^{-W_{ n\to m}/k_{\rm B}T}.\label{eq:expavgwork}
\end{equation}
Finally we can verify the equality by comparing the Eq.~(\ref{eq:expavgwork}) to the $ e^{-\Delta F/k_{\rm B}T}$. Here we will show the experimental realization step by step.
\begin{figure}
\centering
\includegraphics[width=0.8\textwidth]{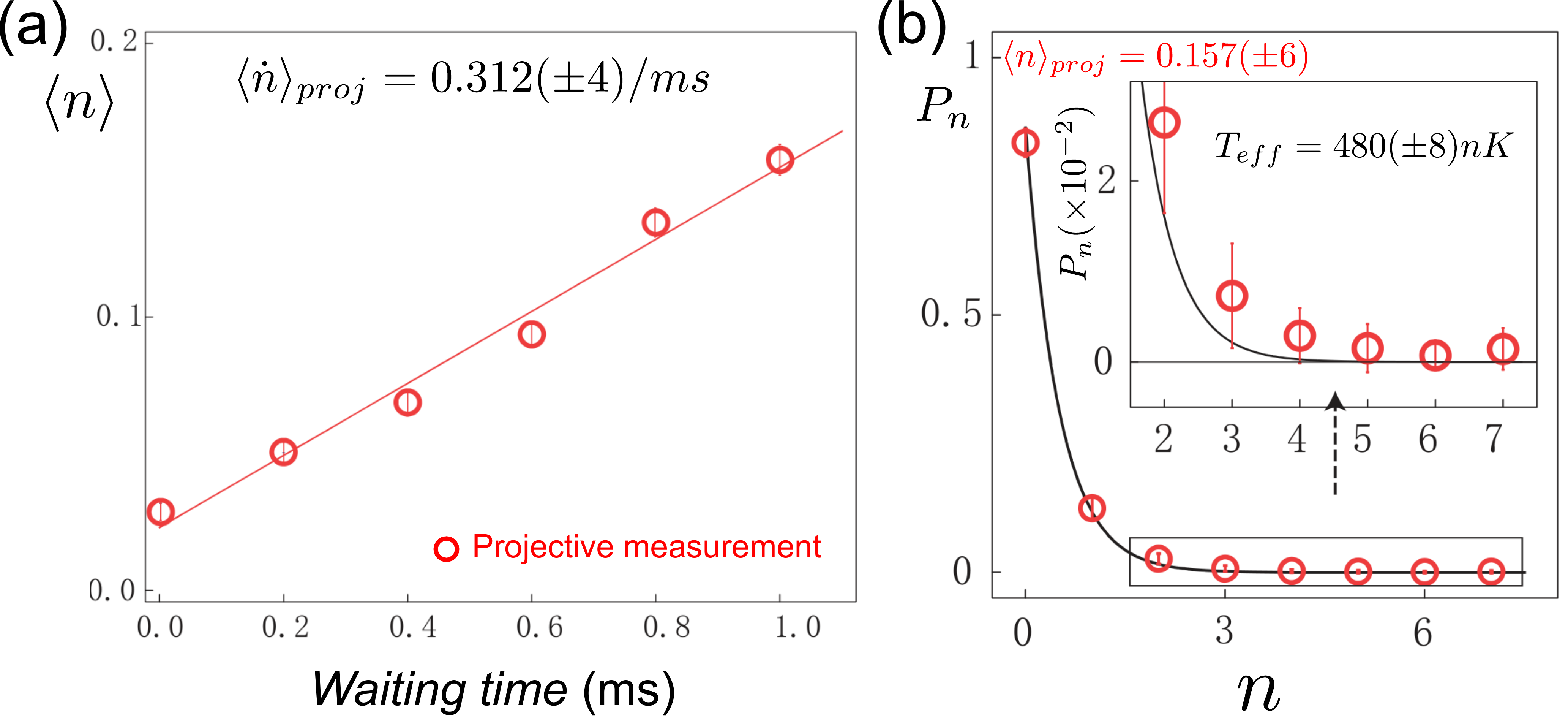}
\caption[Thermal state preparation of the phonon and measurement results.]
{Preparation of the initial thermal state.
(a) The averaged phonon number obtained by the phonon projective measurement as a function of the waiting time. After the side-band cooling, the motional state is cooled near to the ground state and the average phonon number $\avg{n}$ increases linearly with the waiting time, and the slope of the fitted line is defined as the heating rate.
(b) The phonon distribution (circles) measured by projective measurement at the waiting time of 1 ms. The circles are experimental data obtained by the projective measurement and the solid line is a fitted thermal distribution with $P_{ n}^{\rm th} = \avg{n}^{n}/(\avg{n}+1)^{n+1}$. The inset shows the details of the phonon distribution for $n$ lager than 2. The error bars represent one standard deviation of uncertainty.
}
\label{fig:heating}
\end{figure}

\subsection{Experimental Verification in a Closed Quantum System}
In the first example, we utilize the quantum dragged harmonic oscillator model in the previous section~\cite{An15}. The Hamiltonian as the equation (\ref{eq:DH}) can be realized in the interaction picture. In our setup, we use detuning $\nu$ of $2\pi\times 20\ \rm{kHz}$ and the maximum of the Rabi coupling $\Omega(t)$ is $\Omega_{\rm max}=2\pi\times 15\ \rm{kHz}$.

{\it 1. Preparation of the initial thermal state.} The quantum Jarzynski equality requires the thermal state as the initial state. After the standard Doppler cooling and the sideband cooling, the average phonon of the thermal state is about $0.03$. When the ion is trapped in the pseudo-potential generated by the RF field, the ion also feels a DC field fluctuation. It has been theoretically proved that the DC noise plays a role of a high temperature thermal reservoir~\cite{turchette2000heating}. By waiting for $1$ ms, we get a thermal state with average phonon number $\avg{n}=0.157$ $(T_{\rm eff} = 480~nK)$ as shown in Fig.~\ref{fig:heating}.

\begin{figure*}
\centering
\includegraphics[width=1\textwidth]{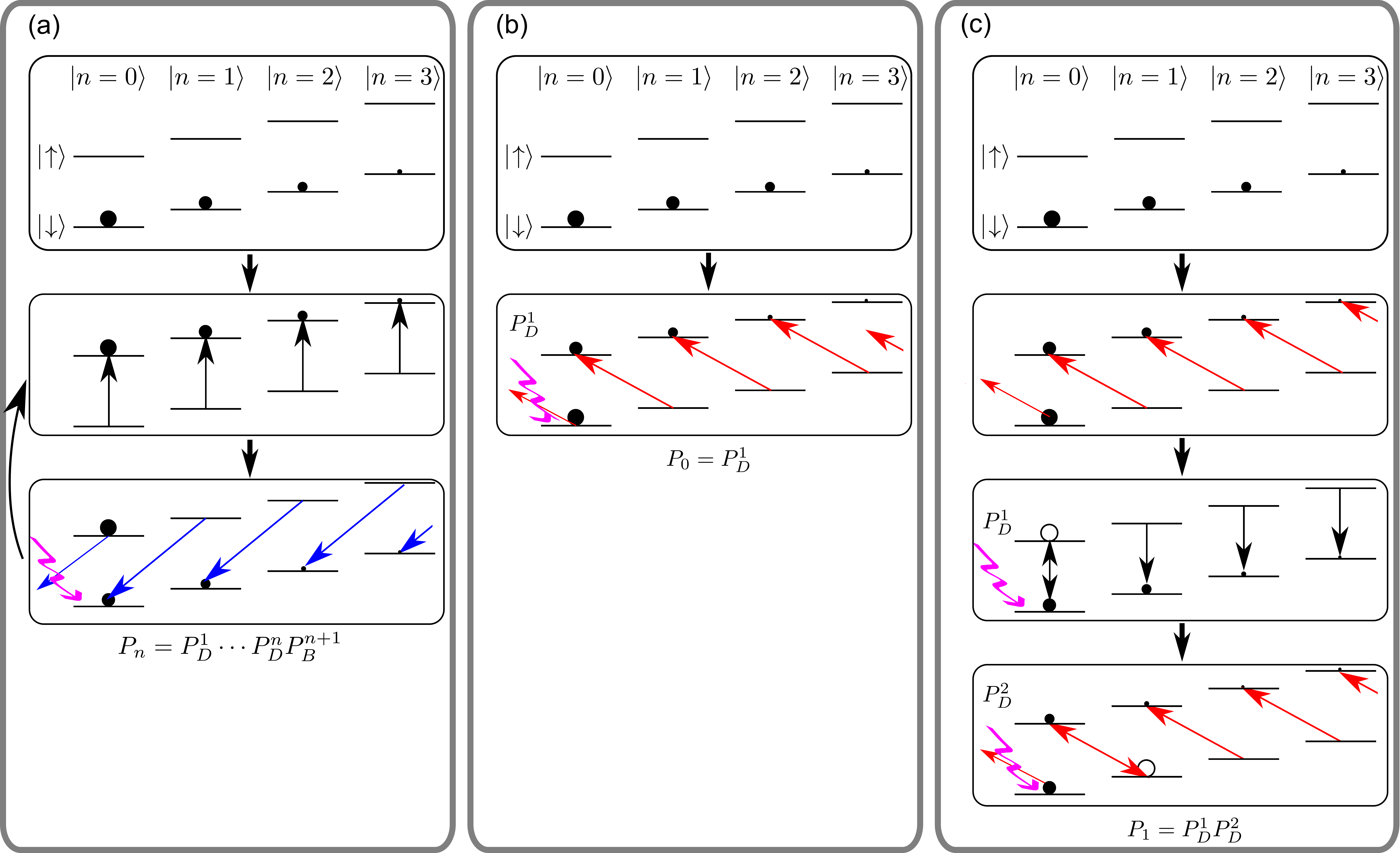}
\caption{
Experimental schemes for the projective measurement of the phonon state.
(a) The fluorescence projective measurement that used in the actual experiments. 
After applying the pulses of carrier and the adiabatic blue-sideband and the sequence of detection, if we see the fluorescence, we know that the state is projected to $\ket{n=0}$. 
If there is no fluorescence, we continue. 
From the number of iterations and the probability that we detect the first fluorescence, we know the projected state and its corresponding probability, respectively.   
(b) The no-fluorescence projective measurement for $\ket{n=0}$. 
After applying the adiabatic red-sideband pulse and the detection sequence, we see no fluorescence with the probability $P_{D}^{1}$, which shows the state is projected to $\ket{n=0}$ with its probability $P_{0}=P_{D}^{1}$.  
(c) The no-fluorescence projective measurement for $\ket{n=1}$. 
We apply the adiabatic red-sideband, the carrier, the detection, the adiabatic red-sideband and the detection.
If the two detections show no fluorescence, the $\ket{n=1}$ state is projected with its probability $P_{1}=P^{1}_{D}P^{2}_{D}$.
The empty circles mean the population has been excluded from the no-fluorescence detection.
For larger $\ket{n}$ states, we can design the similar pulse sequence, which detects  no fluorescence for $n+1$ times successively. 
}
\label{fig:pj}
\end{figure*}

{\it 2. The projective measurement.}  The projective measurement is the key to realize the two-point measurement of the quantum work. The phonon projective measurement we use here was initially proposed for Boson sampling~\cite{shen2014scalable}. The basic operation is shown in Fig.~\ref{fig:pj}.\\
We firstly use the combination of the carrier and the adiabatic blue-sideband pulses to transfer the populations on $\ket{\downarrow,n\geq1}$ to $\ket{\downarrow,n-1}$, and then perform the state-dependent fluorescence detection to interrogate if the system is projected to $\ket{\downarrow, n=0}$. The probability $P^{1}_{B}$ that we see the fluorescence in the first detection sequence is the probability $P_{0}(=P^{1}_{B})$ that the $\ket{n=0}$ state is projected. If we see no fluorescence, we repeat the carrier, the adiabatic blue-sideband and the detection until we see the fluorescence. If we see the first fluorescence at the $(n+1)$th detection, we know the state is projected to $\ket{n}$ with the corresponding probability of $P_{n}=P^{1}_{D}\cdots P^{n}_{D}P_{B}^{n+1}$, where $P^{i}_D$ is the probability of no fluorescence at the $i$-th detection sequence. Note that the internal state is distributed among all Zeeman levels after detecting fluorescence. As a result, after the first fluorescence at the $(n+1)$-th interrogation, we prepare the motional state to the projected number state $\ket{n}$.

An alternative approach to the projective measurement of the phonon state is as follows. We may use pulses of the adiabatic red-sideband and the carrier. To project the $\ket{n=0}$ state, we apply the adiabatic red sideband and then the detection. If we see no fluorescence with the probability $P^{1}_{D}$, the state is projected to the $\ket{n=0}$ state and the corresponding probability is $P_{0}=P^{1}_{D}$. For $\ket{n=1}$, we can use the combination of the adiabatic red-sideband, the carrier, the detection, the adiabatic red-sideband and the detection. If the two detections show no fluorescence, the state is projected to $\ket{n=1}$ with its probability $P_{1}=P^{1}_{D}P^{2}_{D}$ as shown in Fig. \ref{fig:pj}(c). For larger $n$, we can design a similar pulse sequence. With no-fluorescence detection the state is not destroyed by the detection. The disadvantage is that the post-selection process decreases the efficiency of sampling data. Note that in the fluorescence projection method, every result of detection sequence contributes to the sampling data. 

\begin{figure}
\centering
\includegraphics[width=0.4\textwidth]{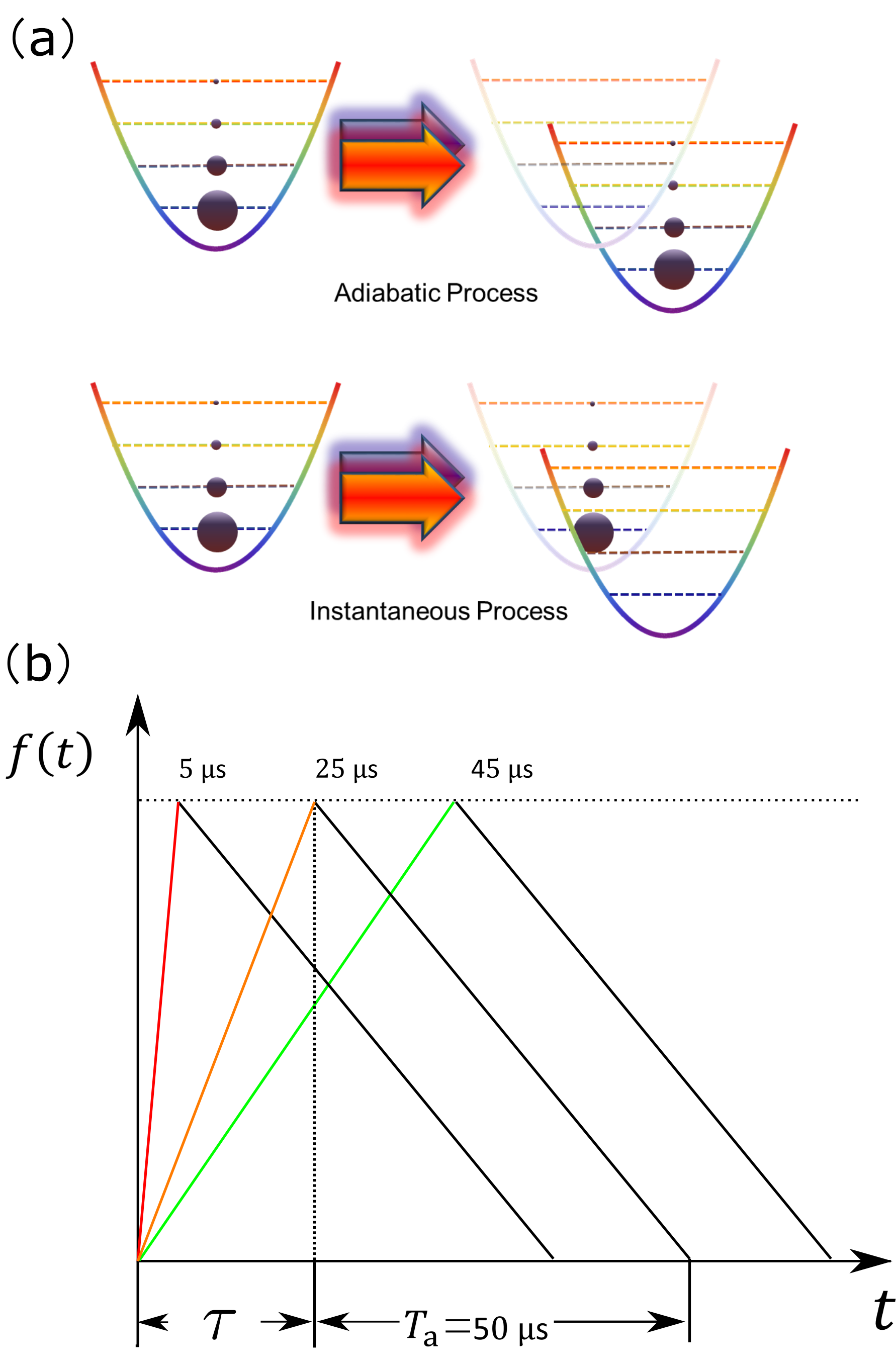}
\caption{The work-performing protocols. (a) The intuitive pictures of the adiabatic and the instantaneous processes for dragging a harmonic potential. The size of the black balls is indicatively proportional to the population probability of the eigenstate (dashed lines). (b) The dragging force as functions of time for protocols of different duration $\tau$. The dragging force is linearly increased to the same maximum value with $t\in[0,\tau]$. These work-performing procedures are followed by an effective adiabatic process with the duration of $T_{\rm a}=50$ $\mu$s (see text), which is one period of the effective harmonic oscillation.}
\label{fig:11}
\end{figure}

{\it 3. Application of work.} After determining and preparing the projected energy eigenstate $|n\rangle_{\rm i}$, we apply work to the system. Specifically, we linearly increase the dragging force to the same maximum value by applying a spin-dependent force on the prepared state for durations $\tau=5$ $\mu$s, 25 $\mu$s and 45 $\mu$s, as shown in Fig.~\ref{fig:11}. The fastest process of 5 $\mu$s duration is regarded as a non-equilibrium process and the longest process with 45 $\mu$s is treated as the adiabatic process.

{\it 4. The second measurement of phonon state.} After the work process, we perform the second energy measurement to get the distribution of quantum work. However, with the dragging force, the final Hamiltonian does not commute with the phonon operator. Thus the energy measurement is not equivalent with the phonon number measurement. Naively, we can adiabatically turn off the laser that generates the spin-dependent force, which would take several tens of the oscillating periods ($2\pi/\nu$) of the effective harmonic oscillator in Eq.~(\ref{eq:DH}) to satisfy the adiabatic criteria. In practice, we find this process is equivalent to an adiabatic process when we reduce the force linearly over a duration of integer multiples of the effective oscillating period.
We can express the time evolution in the form of displacement operator $\hat{D}(\alpha)=\exp(\alpha \hat{a}^{\dagger}-\alpha^{*} \hat{a})$ up to a global phase.
When we increase the laser intensity in $\tau$ and decrease in $T_{\rm a}$ linearly, the displacement in phase space is
\begin{align}
\label{aa} \nonumber
  \alpha=-i\dfrac{\Delta k x_{\rm 0}\Omega_{\rm max}}{2}
\Big(\dfrac{e^{i \tau\nu}(1-i \tau\nu )-1}{\tau\nu^{2}}+\dfrac{e^{i \tau\nu}(1-e^{i T_{\rm a}\nu}+i T_{\rm a}\nu)}{T_{\rm a}\nu^{2}}\Big).
\end{align}
The first term in the parentheses is the contribution from the working process and the second from the transferring process. Here, we can notice that the adiabatic evolution with $T_{\rm a}=+\infty$ is the same for the case with $T_{\rm a}=l\dfrac{2\pi}{\nu}\ (l=1,2,3\dots)$.
When we set $T_{\rm a}=\dfrac{2\pi}{\nu}=50\ \mu s$, we can transfer the phonon distribution of the system Hamiltonian $H_{\rm S}(t=\tau)$ to the distribution of the phonon measurement. 
This fast adiabatic protocol is further developed systematically within the theory of the shortcut to adiabaticity as discussed in Ref.~\cite{An16}. Finally we apply the blue-sideband transition to get the phonon distribution according to Eq.~(\ref{eq:bluedetection}), which is the transition probability $P_{n\to m}$. The experimental dissipated quantum work, $W_{\rm dis}=W-\Delta F$, distribution can be constructed by $P(W_{\rm dis})=\sum_{n,m}P^{\rm th}_{n}P_{n\to m}\delta\left(W_{\rm dis}-\hbar\nu(m-n)\right)$.

\begin{figure}
\centering
\includegraphics[width=0.7\textwidth]{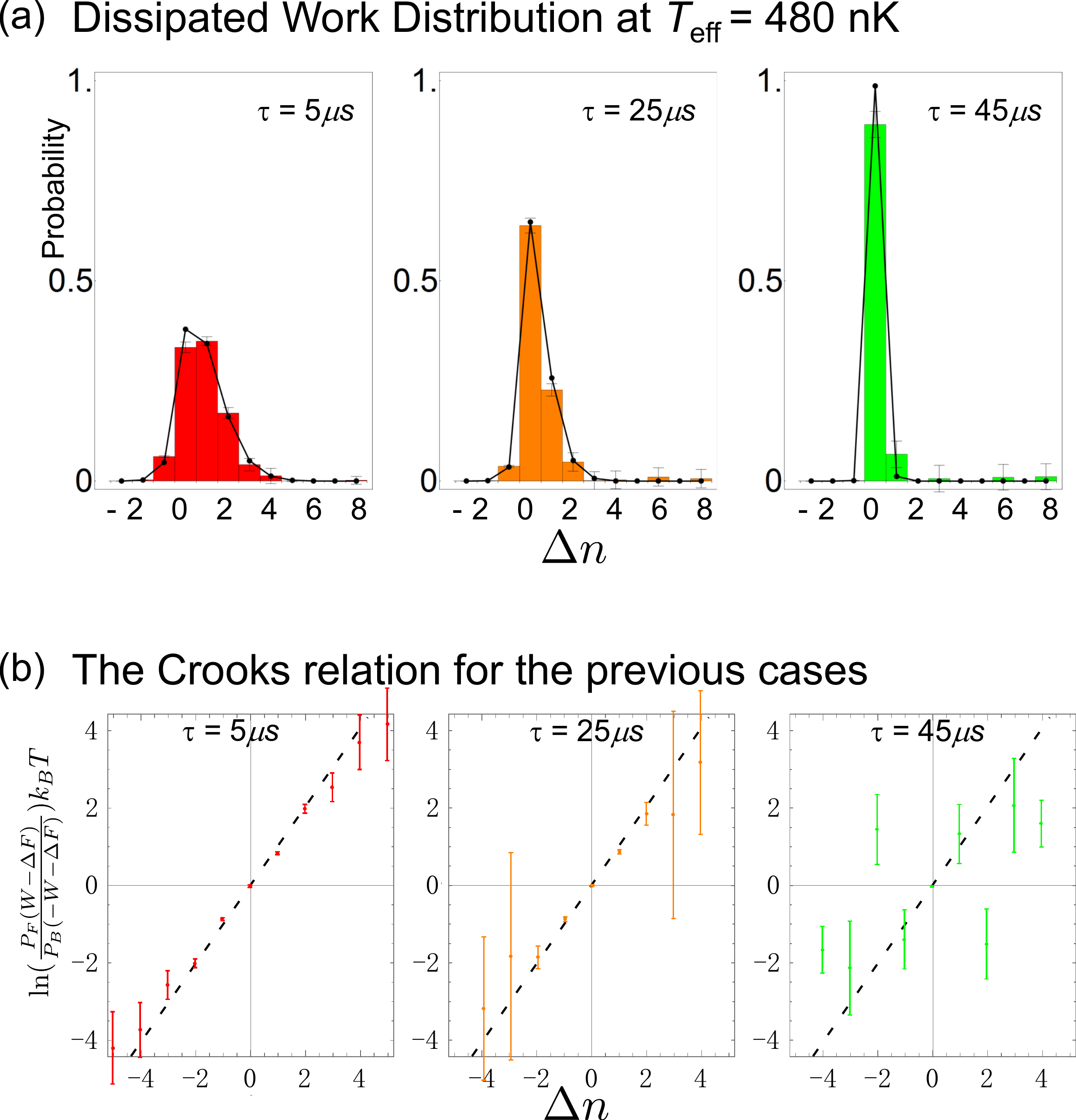}
\caption{The experimental results. (a) For the fastest 5 $\mu s$ working process, we get the widest distribution. 
We can also notice the non-Gaussian feature here. 
For the slowest working process, we get a distribution close to the delta function. 
For all these working processes with different speeds, the exponential average of the dissipated work is one within the experimental error bars. 
(b) The Crooks relation for all the previous cases.
When the fluctuation is relatively large, we can get a good linear relation for the $\tau=5\, \mu$s. When the processes are close to adiabatic, the dissipated work distributions are narrow and the work probabilities for large $|\Delta n|$ become much smaller than the experimental uncertainties, which makes it difficult to clearly observe the Crooks relation.
The dashed lines here are the predictions from the Crooks relation.
Here $P_{F}$ and $P_{B}$ mean the dissipated work distributions for the forward and the backward processes.
}
\label{fig:r}
\end{figure}

The experimentally reconstructed work distributions for the initial thermal distribution with $\langle n\rangle=0.157$ $(T_{\rm eff} = 480~nK)$ are shown in Fig.~\ref{fig:r}(a), based on which the quantum Jarzynski equality is verified within experimental precision. 
It is clear that the ramping of the force with duration $\tau=45 \rm{\mu s}$ is close to an adiabatic process, as there is almost no change in the phonon distribution in the work process. Similar to the results for the classical regime, the mean value and the width of the distribution of the dissipated work increase with the ramping speed. The standard deviations of the dissipated work in the two fast-ramping protocols ($\tau=5\, \rm{\mu s}$ and $\tau=25\, \rm{\mu s}$) show negative dissipated work, as a manifestation of the microscopic ‘violation’ of the second law. Note that in the classical version of our model, only a Gaussian profile of the distribution of dissipated work exists in an open system as well as in an isolated system regardless of the protocol. The non-Gaussian profile at the extremely low initial temperature $\langle n\rangle<1$ is a quantum manifestation. In Fig. \ref{fig:r}(b), we also show the Crooks relations~\cite{Crooks99} for all the three different cases. For the dragged harmonic oscillator model, the backward and the forward processes are identical. We can use the same dissipated work distribution for these forward and backward processes.
We notice for the adiabatic case, the relatively large experimental errors at the large $|\Delta n|$ distorted the linear relation. 

\subsection{Experimental Verification in a Quasi-Open Quantum System}

\begin{figure}[ht]\centering
  \includegraphics[width=0.8\textwidth]{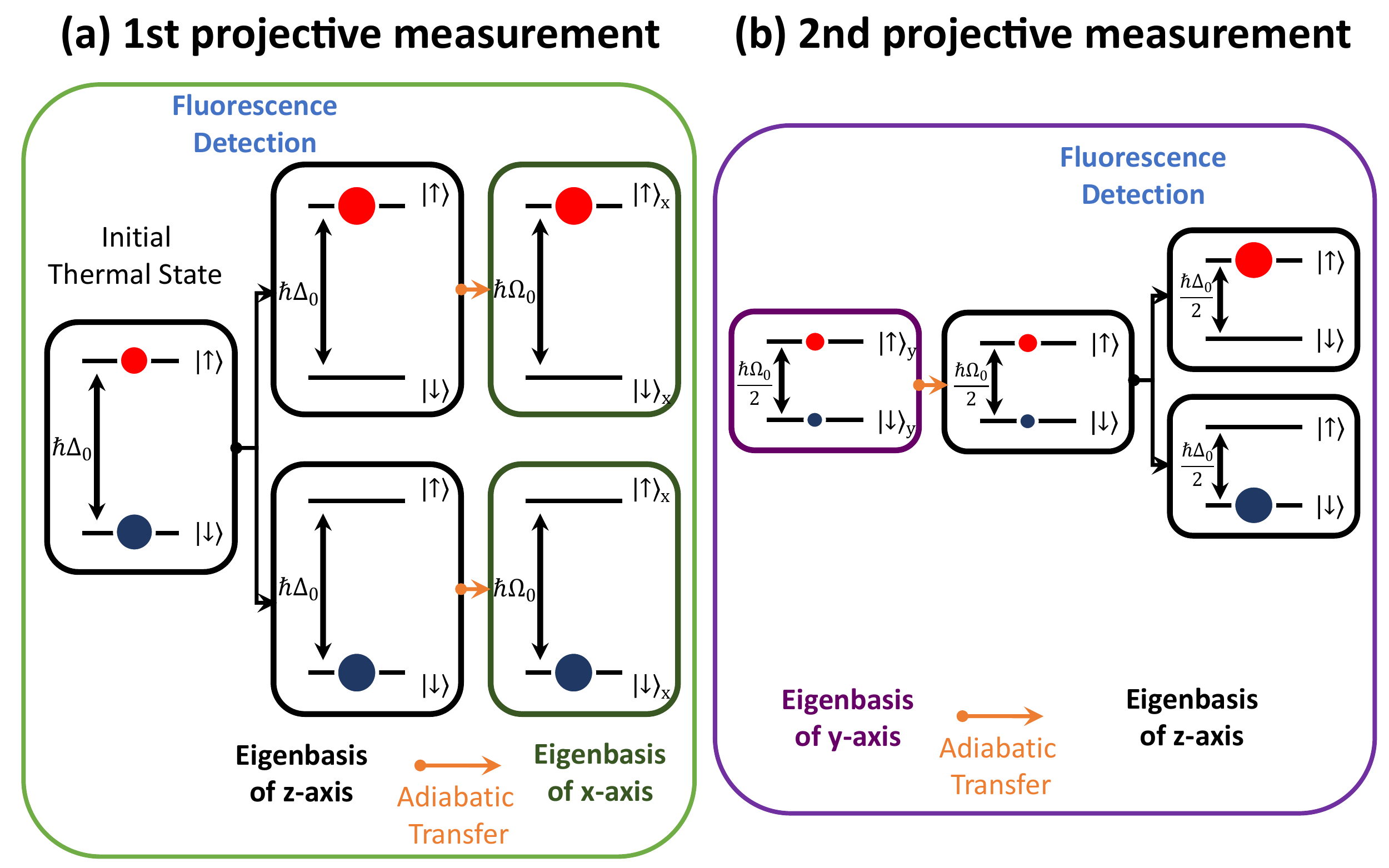}
  \caption{The projective measurements on the internal state. (a) We prepare the initial thermal state and apply the first projective measurement in the basis of $\hat\sigma_{\rm z}$. Then the projected state is transferred to the basis of $\hat\sigma_{\rm x}$ in the adiabatic way. (b) For the second projective measurement, we adiabatically rotate the basis from $\hat\sigma_{\rm y}$ back to $\hat\sigma_{\rm z}$ then apply the state-dependent fluorescence detection to the system.}
  \label{fig:TPM}
\end{figure}

In this example we employ a driven quantum two-level system to verify the quantum Jarzynski equality in the presence of the dephasing effect~\cite{Smith18}. A pair of the internal levels in the ground-state $^{2}S_{1/2}$ manifold, which are $\ket{F=1\,,m_{\rm F}=-1}=\ket{\uparrow}$ and $\ket{F=0\,,m_{\rm F}=0}=\ket{\downarrow}$, are chosen as the two-level system. We modulate both the amplitude and the phase in Eq.~ ($\,$\ref{eq:TLS_drivenH}), leading to a parameter $\tau$-dependent Hamiltonian
\begin{equation}
\hat{H}_{\rm S}(t) = \frac{\hbar\Omega_0}{2}(1 - \frac{t}{2 \tau})[\hat{\sigma}_{\rm x} \cos(\dfrac{\pi t}{2 \tau}) + \hat{\sigma}_{\rm y} \sin(\dfrac{\pi t}{2 \tau})] \label{eq:workprotocol}.
\end{equation}
Here $\Omega_0$ is the energy splitting of the initial Hamiltonian $\hat{H}_{\rm S}(0)$, which is equal to $2\pi \times 50 \, {\rm kHz}$ in the experiment, and the evolution time $t$ varies from $0$ to $\tau$. 

{\it 1. Preparation of the initial thermal state.} 
We utilize the strong decoherence of the Zeeman levels to create an effective thermal state. First we prepare the superposition state $c_\downarrow \ket{\downarrow} + c_\uparrow \ket{\uparrow}$ through optical pumping followed by a resonant microwave transition for a certain duration. Because of the fluctuations of the external magnetic field, this pure state evolves into a mixed state 
$|c_\downarrow|^2 \ket{\downarrow}\bra{\downarrow} + |c_\uparrow|^2 \ket{\uparrow}\bra{\uparrow}$ 
after waiting around ten times the coherence time, which is equivalent to the thermal state with the effective temperature
\begin{equation}
T_{\rm eff} = \dfrac{\hbar \Omega_0}{k_B \ln{(|c_\downarrow|^2/|c_\uparrow|^2)}}.
\end{equation}
In the example of our experiment, the effective temperature is $T_{\rm eff} = 5.63\,{\rm\mu K}$. 

{\it 2. The first projective measurement.} After preparing the effective thermal state in the basis of $\hat{\sigma}_{\rm z}$, we apply the first projective measurement on the internal state. Then this projected state is adiabatically transferred to the basis of the initial Hamiltonian~\cite{Berry09}, shown in Fig.$\,$\ref{fig:TPM}(a). We note that the first projective measurement includes the state-dependent fluorescence detection and the state re-preparation. If the upper state $\ket{\uparrow}$ is detected, the system will be left in a mixed state of all the levels belong to $\ket{^2S_{1/2}\,,F=1}$. In this case we have to re-prepare the upper state using additional optical pumping followed by a microwave $\pi$ pulse before continuing the experiment. Otherwise, if the outcome is $\ket{\downarrow}$, we can directly move to the next step.

{\it 3. Application of work.} Our work protocol is modeled by the Hamiltonian in Eq.~(\ref{eq:workprotocol}), which describes the rotation of the Hamiltonian from the initial $\hat{\sigma}_{\rm x}$ to the final $\hat{\sigma}_{\rm y}$. The parameter $\tau$ is the duration of driving the system. In the experiment we set this parameter to several values of $50\,{\rm \mu s},\ 10\,{\rm \mu s}$ and $5\,{\rm \mu s}$, corresponding to the near adiabatic, the moderate and the fast processes of driving the system. By simultaneously applying an additional stochastic Hamiltonian
\begin{equation}
\hat{H}_{\xi}(t) = \frac{\Omega_0 \xi(t)}{2}[\hat{\sigma}_{\rm x} \cos(\dfrac{\pi t}{2 \tau}) + \hat{\sigma}_{\rm y} \sin(\dfrac{\pi t}{2 \tau})],
\end{equation}
where $\xi(t)$ is the Gaussian white noise, we can simulate the work process undergoing dephasing, which is exactly the same as the evolution described in Eq.~(\ref{eq:dephasing})~\cite{Loreti94,VKampen2007}. Here the dephasing strength $\gamma$ of the Eq.~(\ref{eq:dephasing}) in the basis of the instantaneous Hamiltonian (\ref{eq:workprotocol}) is uniform and proportional to $(\sigma\Omega_0)^2$, where $\sigma$ is defined in Eq.~(\ref{eq:noise}). In our experimental realization, the values of dephasing strength $\gamma$  are set to $0\,{\rm kHz},\ 448\,{\rm kHz}$ and $1340\,{\rm kHz}$ by controlling the noise intensities.

{\it 4. The second projective measurement.} This measurement is similar to the first one. We apply the fluorescence detection after adiabatically transferring state from the $\hat{\sigma}_{\rm y}$ to the $\hat{\sigma}_{\rm z}$ basis, shown in Fig.$\,$\ref{fig:TPM}(b). Because this is the final step of the experimental sequence, the re-preparation of the state is not necessary here.

\begin{figure}[ht]\centering
  \includegraphics[width=0.7\textwidth]{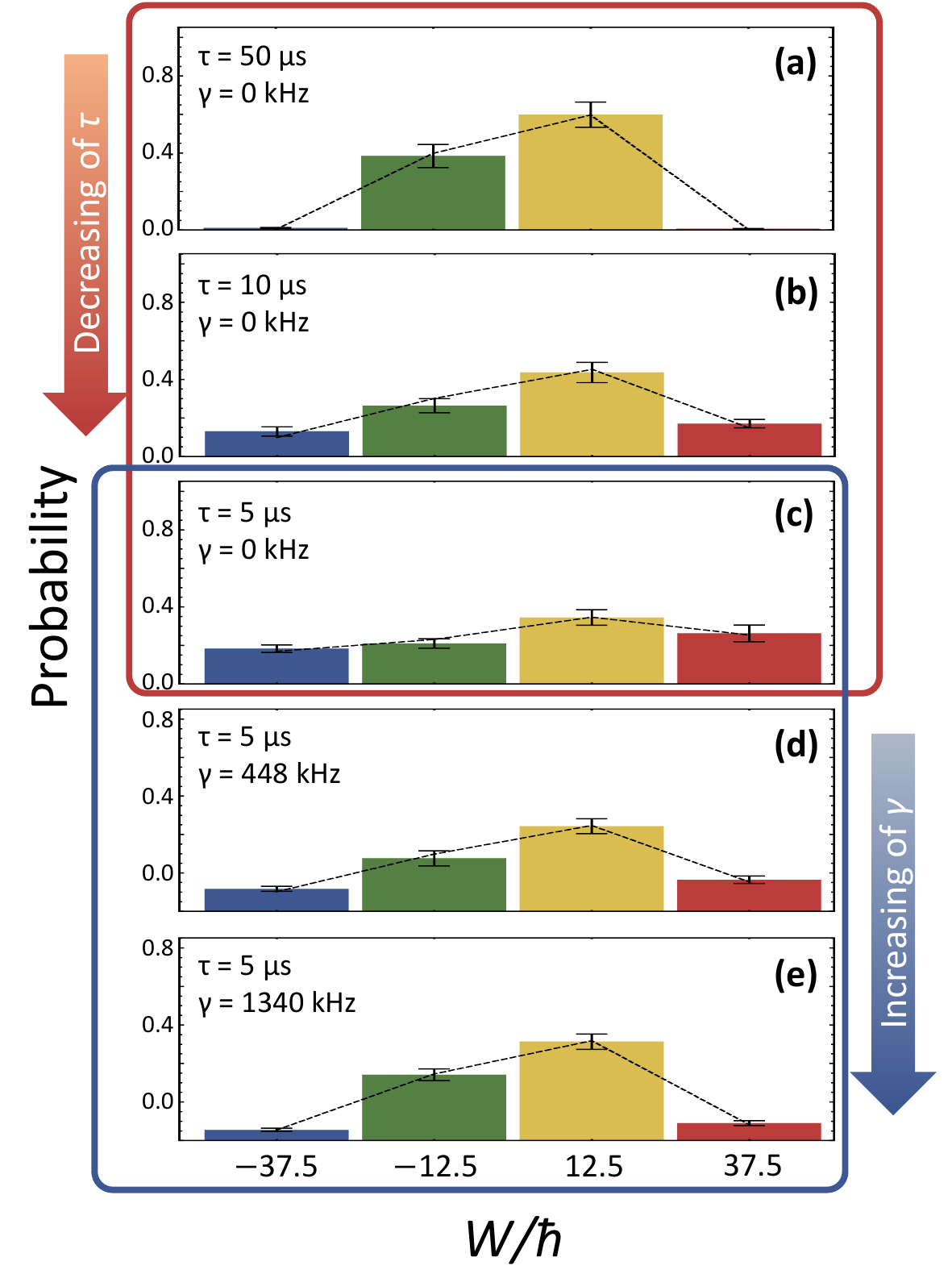}
  \caption{Work distributions for different combinations of the experimental parameters. For (a)-(c) we consider the cases without dephasing, while decreasing the duration of driving system from $50\,{\rm \mu s}$ to $5\,{\rm \mu s}$. For (c)-(e) we keep the duration $\tau$ as 5 $\mu$s, meanwhile increase the strength of dephasing effect from $0\,{\rm kHz}$ to $1340\,{\rm kHz}$.}
\label{fig:workdis}
\end{figure}

We test five combinations of the dephasing strength $\gamma$ and the driving duration $\tau$. The work distributions are summarized in Fig.$\,$\ref{fig:workdis}. It is obvious that the dephasing effect plays a non-trivial role in the process of applying work. The final work distribution depends on the competition between the driving duration $\tau$ and the dephasing strength $\gamma$. Decreasing parameter $\tau$ broadens the work distribution, which means more work is done on the system as shown in Figs.$\,$\ref{fig:workdis}(a-c). However, increasing dephasing strength $\gamma$ tends to narrow the work distribution, and brings it close to the case of the adiabatic process as shown in Figs.$\,$\ref{fig:workdis}(c-e). The experimental results reveal the similarity between the dephasing effect and the continuous measurements on the instantaneous basis of the system, which collapse the wave function of the system and effectively trap the system in one of the instantaneous eigenstates, known as the quantum Zeno effect.

\begin{figure}[ht]\centering
  \includegraphics[width=0.7\textwidth]{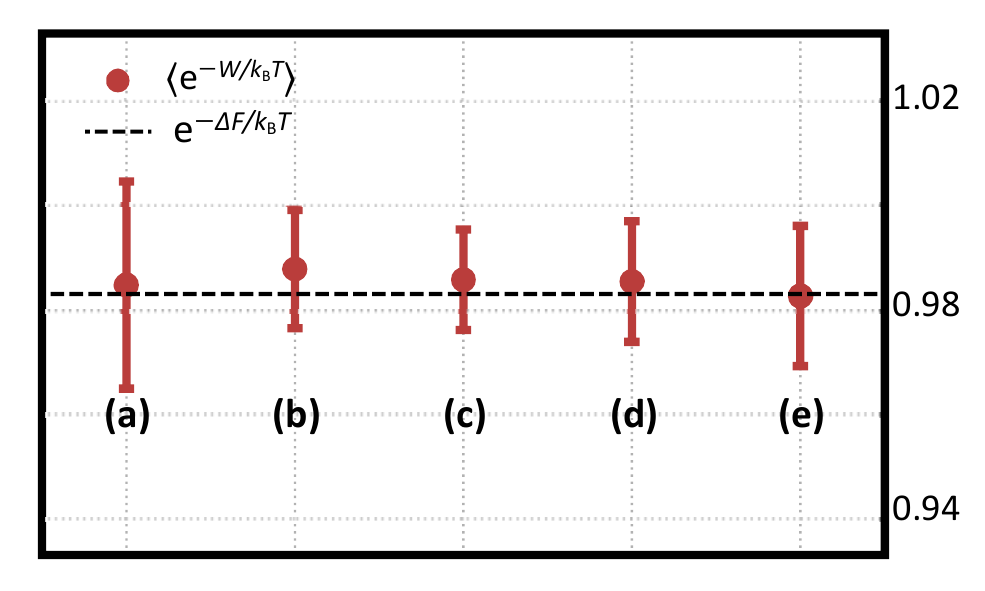}
  \caption{Comparison between the exponential average of work for distributions ((a)-(e) shown in Fig.$\,$\ref{fig:workdis}) and the exponential of the calculated free energy difference.}\label{JE} 
\end{figure}

Meanwhile, the quantum Jarzynski equality can be verified with all the work distributions as shown in Fig.$\,$\ref{JE}. The  quantity $\langle e^{-W/k_{\rm B}T}\rangle$ in the left hand of Eq.~(\ref{eq:JarzynskiEq}) is obtained by using the experimental work distribution as discussed in Eq.~(\ref{eq:expavgwork}), while the right hand of Eq.~(\ref{eq:JarzynskiEq}) was calculated by the theoretical difference of the free energy between $\hat{H}_{\rm S}(\tau)$ and $\hat{H}_{\rm S}(0)$. The experimental results are consistent with the theoretical predictions within the experimental errors.

\section{Conclusion and Discussions} \label{sec:discussions}
In this chapter, we present how to employ a single ion trapped in a harmonic potential for the experimental test of the quantum Jarzynski equality. Using projective measurements of the motional mode, we verify the equality for a closed quantum system. With the internal levels and properly designed noise, we perform the experimental test of the equality for a quasi-open quantum system. Together with further theoretical developments, these demonstrations would be important stepping stones to fully confirm the equality and the fluctuation theorem for open quantum systems~\cite{Breuer2002,Pigeon2016}. Besides being used in verifying the quantum Jarzynski equality, our experimental breakthrough could be applied to realize many other thought experiments in quantum thermodynamics. For example, it could be applied to the studies of
quantum heat engines by experimentally exploring the relation between work and heat in thermodynamic cycles~\cite{Quan07,Lutz12,rossnagel2016single,maslennikov2017quantum}. 

\bigskip

\acknowledgements

ACKNOWLEDGEMENTS

This work was supported by the National Key Research and Development Program of China under Grants No. 2016YFA0301900 and No. 2016YFA0301901 and the National Natural Science Foundation of China Grants No. 11374178, No. 11574002, and No. 11504197.

%

\end{document}